# Performance Evaluation of Checkpoint/Restart Techniques

For MPI Applications on Amazon Cloud

Basma Abdel Azeem
College of Computing and Information Technology
Arab Academy for Science, Technology and Maritime Transport
Cairo, Egypt
basmaabdelazeem@hotmail.com

Manal Helal
College of Engineering and Technology
Arab Academy for Science, Technology and Maritime Transport
Cairo, Egypt
manal.helal@gmail.com

*Abstract*—Distributed applications running on a large cluster environment, such as the cloud instances will have shorter execution time. However, the application might suffer from sudden termination due to unpredicted computing node failures, thus loosing the whole computation. Checkpoint/restart is a fault tolerance technique used to solve this problem. In this work we evaluated the performance of two of the most commonly used checkpoint/restart techniques (Distributed Multithreaded Checkpointing (DMTCP) and Berkeley Lab Checkpoint/Restart library (BLCR) integrated into the OpenMPI framework). We aimed to test their validity and evaluate their performance in both local and Amazon Elastic Compute Cloud (EC2) environments. The experiments were conducted on Amazon EC2 as a well-known proprietary cloud computing service provider. Results obtained were reported and compared to evaluate checkpoint and restart time values, data scalability and compute processes scalability. The findings proved that DMTCP performs better than BLCR for checkpoint and restart speed, data scalability and compute processes scalability experiments.

I. INTRODUCTION

Cloud computing is a large-scale distributed computing paradigm, which is based on virtualization and on demand dynamic scaling. It is managed by various providers to reduce establishment and management costs by supporting high powerful software and hardware computing resources to application programmers. Such resources were previously available only to large technology centres and national laboratories [1]. Migrating legacy-distributed applications to the cloud infrastructure for large-scale deployment requires planning for nodes failure possibilities. Message Passing Interface (MPI) [2] is the most widely used programming model in distributed parallel environment. A set of MPI program tasks utilise their own local memory during computation, And communicate with one another by sending and receiving messages. MPI is specifically used to aid parallel execution of applications over a network of separate computers. MPI programs generally run the same piece of code on all of the targeted processors.

As the computing environments were getting larger, a higher number of different types of failures started to appear. These failures should be fully tolerated by software systems especially MPI programs to avoid the cost of repeating the work done before the failure occurred. Several fault tolerance techniques have been proposed to handle this problem. Checkpoint/restart is one of the most widely adopted rollback recovery class of these techniques. It allows an application's state to be preserved to a stable storage device and recovered at a later time. The probability of the different faults occurrences increases in proportion to the time required by the application to complete, and also in proportion to the number of nodes it is using. Checkpoint/restart is needed in parallel applications to help in exploring the reasons of any possible error by having a specified roll back snapshots in certain points of the long computation time and to make it safer to extend the parallel width of software jobs in the future [3, 4].

In this paper, we chose two from the most significant and most widely used checkpoint /restart implementations to ensure their applicability to be used on Amazon Elastic Compute Cloud (EC2) and evaluate their performance in local environment and on Amazon EC2. This is to ensure that legacy parallel applications using these fault tolerance techniques can work properly on Amazon EC2. We also investigated checkpoint and restart time values and some scaling issues related to the checkpoint/restart behaviour of these tools within Amazon EC2 environment. The evaluated checkpoint/restart tools are: 1) the kernel based Berkeley Lab Checkpoint/Restart library (BLCR) [5, 6], which is integrated into the OpenMPI [3,7] framework as its fault tolerance support; and 2) Distributed Multithreaded Checkpointing (DMTCP) [8]. We focused on providing checkpoint/restart facilities for parallel applications that communicate using MPI. A brief background concerning the selected techniques is provided in section III. We used Amazon EC2 [9, 10] as the cloud environment in our experiment. Amazon as a commercial cloud provider offers a suitable environment to test and run different types of applications in the cloud. It is a virtual resources environment that offers launching and managing virtual machine (VM) instances through a web services application-programming interface (API). Amazon EC2 supports various instance types that differ in performance characteristics.

This paper is organized as follows: section II describes the literature review of the design specifications related to checkpoint/restart tools implementation. Section III highlights some specific characteristics of the chosen tools to help understanding their behaviour. Section IV explains the experiments procedure and used methods. Section v discusses and visualizes the obtained results. Section VI presents the analyses and conclusions. Finally the future work is briefly discussed in VII.

## II. LITERATURE REVIEW

Management of various failures in high performance computing environments and clouds has been studied previously in many research endeavours including [11] and [12]. The checkpointing and rollback recovery techniques have been widely used to support seamless fault tolerance for MPI programs in distributed systems as well as typical desktop application. The work presented in [13] is an example of an effort to classify fault tolerant MPI systems according to the used technique and the software stack level at which the fault-tolerant message passing systems is managed. A discussion of different checkpoint/restart packages and a summary of some of their features are found in a number of surveys such as [14, 15, 16, 17, 18, 19, 20]. Most of these surveys adopted a theoretical trend. Our work focused on experimentally understanding the different aspects of the design of the selected checkpoint/restart tools and their effect on performance while changing some of the computation factors. The findings presented in the analysis and conclusions section can help legacy MPI distributed application developers to choose the suitable checkpoint/restart tool while migrating their work to the cloud.

### A. Checkpoint/Restart General Requirements

Many checkpoint/restart implementations are available. These implementations differ in many aspects including how each process state is preserved, how much of each process state is preserved, how the state is stored and how it deals with APIs and command line interfaces. Several factors should be considered to assess various checkpoint/restart solutions [8, 15, 21]. A good implementation should try to reduce storage, cost and computation overhead; allow for an on demand and autonomic checkpointing; should not require source code modifications, recompilation or relinking with other binaries; can do its job for a wide variety of applications; and should not be tied to certain versions of an application or operating system [14].

### B. Checkpoint/Restart for Distributed Execution

Three main approaches are found to checkpoint and restart a distributed application: coordinated checkpointing, uncoordinated checkpointing associated with message logging, and Communication Induced Checkpointing (CIC). A detailed analysis of these approaches can be found in [4].

The most widely used approach is coordinated checkpointing. This approach stores a snapshot of all processes in the distributed system, such that the global application checkpoint is guaranteed to be consistent. This can be achieved by an orchestrated cooperation among all participating processes in the determination of their individual local checkpoints [22, 23]. The resulting set of local checkpoints constitutes the global checkpoint, which is said to be consistent if it's contained group of local checkpoints forms a consistent global state of the application. In the context of MPI applications, a consistent global state is a combination of the states of all individual processes and the states of all communication channels at any moment during a correct and failure-free execution of a distributed computation. Such a consistent state can be used to restart application execution correctly upon failure [24, 25]. Some protocols have been introduced to achieve global coordination. For example a non-blocking checkpointing coordination protocol [26] attempts to stop running applications that may affect the coordinated checkpointing consistency. The use of synchronized clocks for checkpointing is also preferred [14]. Some previously introduced algorithms like The Connection Machine (CM5) [26] stores the in-transit messages while storing the whole state of all routers in the network. Other algorithms like the well-known Chandy and Lamport [22] attempts to empty the network from any in-transit messages such that it prevents any process from sending any messages and wait to ensure that all messages that were already sent have reached their destinations and save them with the process snapshot. Each participating process stores a single checkpoint so the storage overhead is significantly reduced. Using this approach it is simple to implement the recovery process. If failure occurs, all processes are forced to roll back to the most recent checkpoint even when only one single process has a failure and the execution continues from that point. In order to have this simplified recovery, coordination protocols consume more time in computing the application global checkpoint before writing it to the permanent storage thereby increased execution and time overhead [4]. However, most scientific programs are naturally iterative allowing the checkpoint computing to be performed while progressing from iteration to the next. This also guarantees a minimum checkpoint size [27]. DMTCP [8] and BLCR [5, 6] are both examples of coordinated checkpoint techniques.

In uncoordinated checkpointing all participating processes checkpoints are independent from each other. These techniques generally rely on logging messages and possibly their temporal ordering for asynchronous checkpointing [28]. On a theoretical basis it is considered that messages reception by a process within a distributed computation is the only factor that will determine the process execution, thus all consequent state changes. So message-logging techniques and process checkpointing techniques are both able to completely describe the execution state of any process. If failure occurs, the set of saved checkpoints will be searched for a global consistent state from which application can continue execution properly [25]. MPI over Chameleon for Volatile resources (MPICH-V) [13] is an example of uncoordinated checkpoint techniques. The main advantage of uncoordinated checkpointing is that a checkpoint can be taken at the most convenient time for each process thereby reducing overhead. To be more efficient, a process can perform checkpoints when its state is small [27]. Each process needs to maintain multiple checkpoints, which increases the storage overhead [24].

Using uncoordinated checkpointing, it may be difficult to obtain a global consistent state. Consequently the checkpoint will be ineffective. Therefore, uncoordinated checkpointing is prone to the domino effect [29], which is caused by rollback propagation from one process to another. This might continue until it reaches the beginning of the execution and may lead to undesirable loss of some or all of the computational work [14].

CIC techniques combine some aspects of uncoordinated and coordinated techniques. A CIC technique piggybacks the message reception orders on all application messages causally. This helps to guarantee an overall order of messages delivery. The CIC technique enforces some chosen processes to be checkpointed, when it suspects having an inconsistent state, to preserve the recovery line progress. Thus each process can select its most suitable time to be checkpointed; for example when its process state is small, this reduces the saving overhead. CIC protocols can avoid domino effect in uncoordinated checkpointing. However, it turns out to be impractical because of its considerable impact on network communication and application performance due to the management of the piggybacked data. It does not scale well with increasing number of processes, which will lead to an increasing number of forced checkpoints and increasing storage requirements [30].

*C. Checkpoint/Restart Implementation Level*

Checkpoint/restart systems need to be integrated with the application and the operating system. This can be achieved at any of these three levels: system level, user level or application level. In case of system level implementation, checkpoint/restart procedures may be included in the OS kernel usually as a dynamically loaded kernel module. It is always transparent to the user since no changes are required to the programs to make them ready for checkpointing. Kernel level implementations are not portable to other platforms and require the kernel source code to be available for modification, which is not always possible. With user level implementations: a user level library provides the checkpoint/restart functionality. Thus applications should be linked to this library. Because of this necessary explicit linking, this approach is usually not transparent to users. It depends on the implementation details. One drawback of this approach is that the library should be permitted to make some system calls to access system data and this is not always available. In application level implementations, all checkpoint/restart activities are written into the application or directly injected into the application code using an automated pre-processor; thus these implementations are not transparent. It provides more control over the checkpoint/restart processes; but requires very deep understanding of application details [14].

### III. SELECTED TECHNIQUES CHARACTERISTICS

Berkeley Lab Checkpoint/Restart library (BLCR) is a transparent system-level kernel based infrastructure for checkpoint/restart. It is an open source checkpoint/restart library that is deployed on several distributed systems. It stores checkpoint data: (stack, heap, registers, signals, etc.). BLCR can checkpoint/restart a single node as a standalone system. It can also be used to checkpoint/restart parallel applications running on more than one node by adding it as an additional configuration to a parallel communication library or to a scheduling system. BLCR supports serial and multithreaded applications. It works on x86 and x86_64 (Opteron/EM64T) systems having Linux kernel from 2.6.x through 3.7.1. It does not checkpoint or restart open files or sockets like Transmission Control Protocol/User Datagram Protocol (TCP/UDP). [5, 6] The OpenMPI framework has integrated BLCR. This produced an infrastructure that supports distributed coordinated checkpoint/restart fault tolerance within Open MPI's modular component architecture. The resulting framework supports the BLCR and SELF-systems for checkpoint/restart. The aim from having the SELF component is to make checkpoint\restart at the application level feasible through supporting the program callbacks required to checkpoint and resume involved operations [3, 7].

DMTCP is a freely available user-level coordinated checkpoint/restart library implementation for distributed computations. It supports sequential and multi-threaded computations across single/multiple hosts. DMTCP is transparent since it doesn't require any recompilation or re-linking to do its job. It entirely works in the user area so there are no added kernel modules and no need for root privileges. A dynamically injected library and a checkpointing manager thread are spawned in each application process. To save the program state, DMTCP stores user space memory, processor state, kernel state, and data in the network. It is demonstrated that DMTCP can checkpoint/restart a wide range of well-known applications including Scripting languages like MATLAB, Python, PHP and Ruby and distributed platforms like MPICH2 and OpenMPI. Checkpoint/restart can be scheduled to run periodically or it can be manually initiated. DMTCP runs on most Linux distributions, and supports both x86 and x86_64 (Intel/AMD for 32- and 64-bits), and 32-bit ARM (ARMv7). It supports Linux 2.6.9 and later [8].

Beside their popularity, the two selected tools were specifically chosen because of many reasons [5, 6, 8] including:

- Their broad application coverage across a large array of both ordinary desktop applications and high performance scientific applications.

- Their ability to work under many Linux flavours and kernel versions.

- Their support of the critical features of transparency.

- Their portability across different Central Processing Unit (CPU) architectures.

- Their ability to handle distributed, multithreaded cluster computations.

- In addition, DMTCP was chosen because of many other advantages. It is a general lightweight solution for socket-based distributed applications, which can also checkpoint high performance networks, such as InfiniBand (IB) [31]. It has some facilities that could be beneficial, for example if the user needs more control over the checkpoint process, an extra library could be added to allow the application to delay checkpointing till a critical part of code finish its execution. Another helpful feature of DMTCP is that it is consisting of two separate layers. The first layer handles checkpoint details of processes running across several networked nodes, so it copies the inter-process data to the user space. The second layer is a single process checkpointer MTCP [32]. This two-layer architecture of DMTCP can help it to work in non-Linux environments by replacing the single process checkpointer with any valid package that can work under the new environment [8].

- BLCR in its turn has attractive features that make it a good candidate to be selected for this experiment. BLCR is particularly notable because of its widespread usage. Libraries or applications can easily integrate with BLCR because it has a very simple interface. Many MPI libraries (including some versions of OpenMPI, Local Area Multicomputer /Message Passing Interface (LAM/MPI) [24], MVAPICH2 [33], and MPICH-V [13]) are integrated with BLCR to provide distributed checkpointing. BLCR also can register call back function that was written at the user level in such a way that it can be called when there is a need to take a checkpoint or to perform restarting [5, 6].

## IV. EXPERIMENTAL SETUP

The selected two checkpoint/restart tools are: 1) BLCR integrated into OpenMPI framework as its fault tolerance support; 2) and DMTCP. The conducted experiments aimed to ensure that the use of the selected checkpoint/restart tool is valid on Amazon EC2 and evaluate their performance specifically while increasing data size, participating processes, and number of instances as required by the application. Experiments used different data sizes and were conducted on both a single multicore node and on an Amazon EC2 instances Linux cluster. The single node tests were made on two different environments: the first one is a local machine that has 64-bit platform, 4 GB memory, and Intel(R) core(TM) i5 CPU M 430 @ 2.27 GHz processor; and the second machine is a single Amazon EC2 instance. The type of this instance is the compute optimised c3.xlarge instance type, which has 64-bit platform, 7.5 GB memory, and 4 virtual CPU (vCPU). Each vCPU on C3 instance is a hardware hyper-thread from a 2.8 GHz Intel Xeon E5-2680v2 (Ivy Bridge) processor. Amazon EC2 compute-optimised instances are recommended for highly consuming compute power applications like parallel programming, complex science and engineering problems [34]. This single Amazon EC2 instance was used to test checkpoint / restart performance of both tools within Amazon EC2 environment including normal and forked DMTCP checkpointing. In forked checkpointing, a child process is forked, this child process writes the checkpoint image, while the parent process continues to execute [8]. The Amazon EC2 instances Linux cluster we used is comprised of 4 compute nodes. Each is one of the Compute optimised c1.xlarge instance type, which is 64-bit platform and has 7 GB memory and 8 virtual CPU (vCPU). The high-CPU c1.xlarge instances run on systems with dual-socket Intel Xeon E5410 2.33GHz processors [35]. The nodes are interconnected by 10 Gigabit Ethernet network. To simplify and automate configuring, building, and managing our cluster of virtual machines on Amazon EC2 cloud, the open source StarCluster cluster-computing toolkit for Amazon EC2 [36] has been utilised. StarCluster can help any user to easily build a cluster environment from the cloud compute nodes (instances) that can be used for parallel and distributed applications.

The checkpoint files created by any of the checkpointing tools should be protected against failures. For Amazon EC2 instances environment, checkpoint files are stored on the local partition of the VM. These files can be stored on Amazon Elastic Block Store (EBS) volumes to increase the availability and durability. EBS is a block based storage service offered by Amazon. It provides automatically replicated persistent storage volumes that can be attached to an Amazon EC2 instance while it is in the running state. Each EBS volume size currently ranges from 1 GB to 1 TB. Having used StarCluster toolkit and images, NFS is configured on EBS volumes. Files saved in one instance EBS volume with NFS configured properly, is automatically replicated to the other instances. Therefore a failure of one instance will not erase all replicas of its checkpoint files on the other instances of the Amazon EC2 cluster created by StarCluster. For additional reliability boost, EBS volume can be backed up to the highly reliable object based Amazon Simple Storage Service (S3) by creating a snapshot of that volume. S3 file system replicates data across multiple geographically dispersed Amazon's data centre. So in case of failure in one data centre, data still can be reached through other replicas in different places [37, 38]. Offering EBS and S3 fault tolerance services by Amazon was a good reason to consider Amazon as a good candidate for this experiment, which examines tools and environments that can offer full protection against failures for the running applications.

The OS used is Linux Ubuntu 12.10, 64-bit platform. We used OpenMPI 1.6.5 which is configured to use BLCR 0.8.5 for fault tolerance. The DMTCP used was version 1.8.5. For all the conducted experiments we used the MPI version of the NASA Advanced Supercomputing (NAS) product, which is called NAS Parallel Benchmarks (NPB) suite [39]. NPB is a suite of programs used to evaluate the performance of parallel systems. We used Lower-Upper Gauss-Seidel solver (LU) and Block Tri-diagonal solver (BT) benchmarks of the NPB of both classes A and B. LU and BT are the most communication intensive applications among NPB [40]. This feature was a good motivation to select these two benchmarks for this experiment. It can help to test the effect of large number of in transit messages that will be produced on the performance of the examined checkpoint/restart tools. According to NPB documentation [41], classes A and B are considered suitable for standard test problems and there is a significant problem size increase and parameters change going from one class to the next. Problem sizes and parameters for class A and B for the selected benchmarks are detailed in table I [41].

The 9th International Conference on INFOrmatics and Systems (INFOS2014) – 15-17 December
Parallel and Distributed Computing Track

TABLE I. PROBLEM SIZES AND PARAMETERS FOR LU AND BT BENCHMARKS FOR CLASSES A AND B DEFINED IN NPB 3.3

| Benchmark | Parameter | Class A | Class B |
|---|---|---|---|
| BT | Grid size | 64 x 64 x 64 | 102 x 102 x 102 |
| | No. of iterations | 200 | 200 |
| | Time step | 0.0008 | 0.0003 |
| LU | Grid size | 64 x 64 x 64 | 102 x 102 x 102 |
| | No. of iterations | 250 | 250 |
| | Time step | 2.0 | 2.0 |

## V. RESULTS

Experiments were conducted to assess both checkpoint time and restart time using the two selected checkpoint/restart tools. In order to test data scalability, NAS parallel benchmarks LU and BT of both classes A and B were used. Benchmarks for both classes C and D could not be used because of their large memory requirements, which exceed our test budget. To test computing nodes scalability, the experiments were conducted using varying number of nodes ranging from single Amazon EC2 instance to 4 instances Amazon EC2 cluster. We created a batch file that runs the application, then checkpoints it, and then restarts it. The restart process was performed on the same machine - or the same virtual machine in case of Amazon EC2 - in which we took the checkpoints. Figure 1 and Figure 2 show the checkpoint time and restart time respectively at varied application scales resulted from the first test performed on a local single node. Figure 3 and Figure 4 show the checkpoint time and restart time obtained when running tests on a single Amazon EC2 instance. Figure 3 shows the checkpoint time of the only test we did using forked checkpoint facility [8] offered by DMTCP. This facility will be discussed in section VI. Figure 4 shows the restart time. While Figure 5 and Figure 6 show these times when running tests on a 4 compute nodes Amazon EC2 cluster.

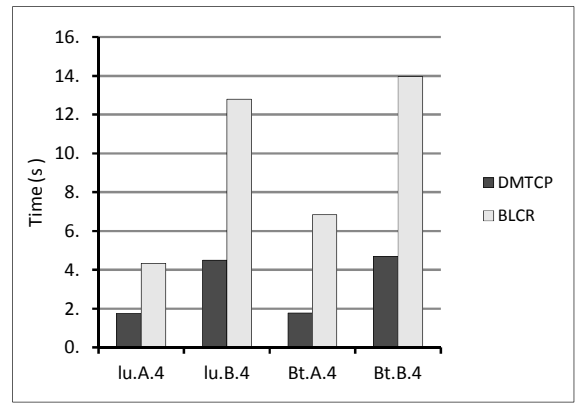

Fig. 2. Local single node restart time.

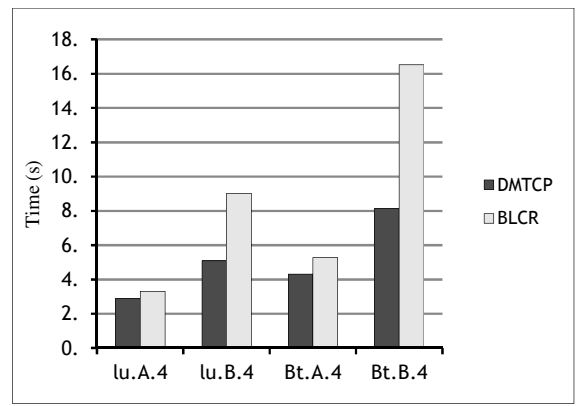

Fig. 3. Single Amazon EC2 node checkpoint time using forked checkpointing for DMTCP.

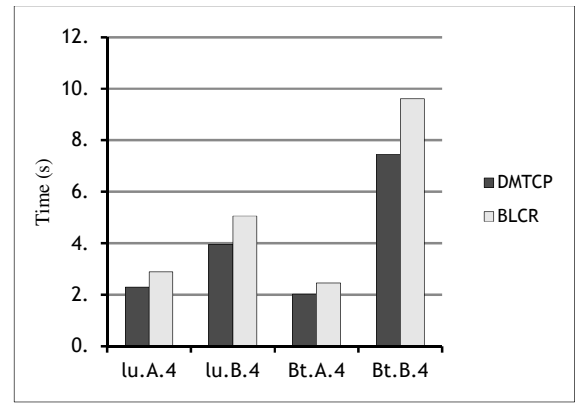

Fig. 4. Single Amazon EC2 node restart time.

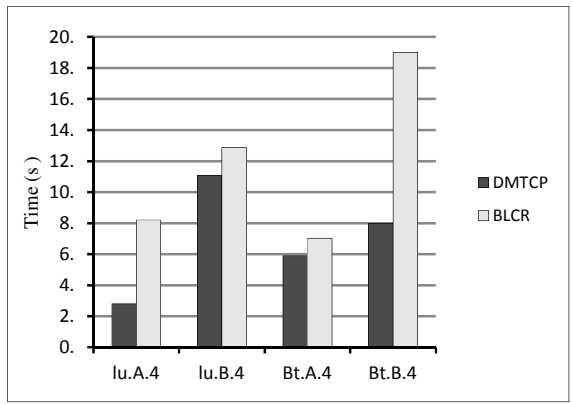

Fig. 1. Local single node checkpoint time.

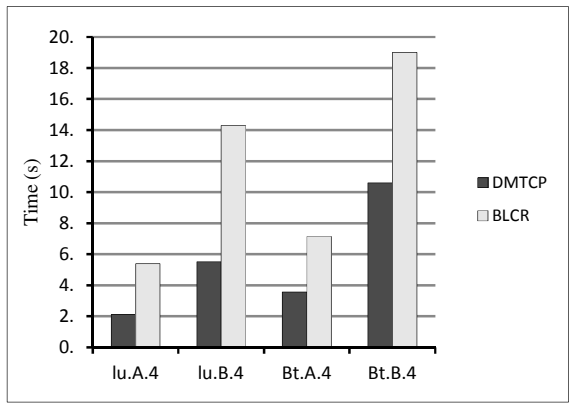

Fig. 5. 4-nodes Amazon EC2 cluster checkpoint time.

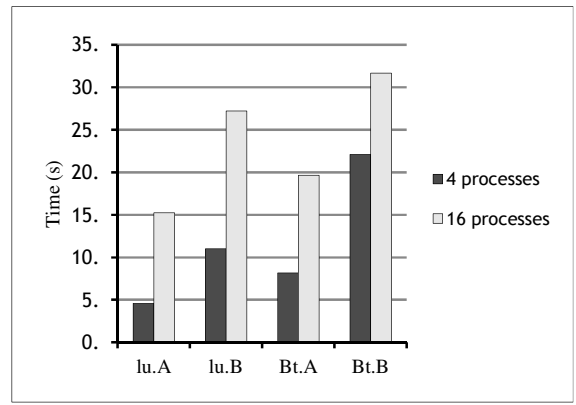

Fig. 7. Single Amazon EC2 node checkpoint time (DMTCP).

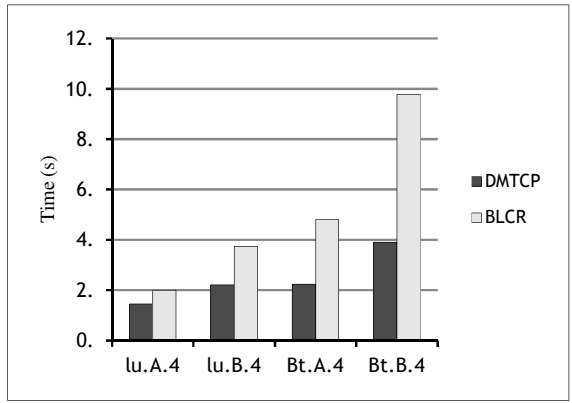

Fig. 6. 4-nodes Amazon EC2 cluster restart time.

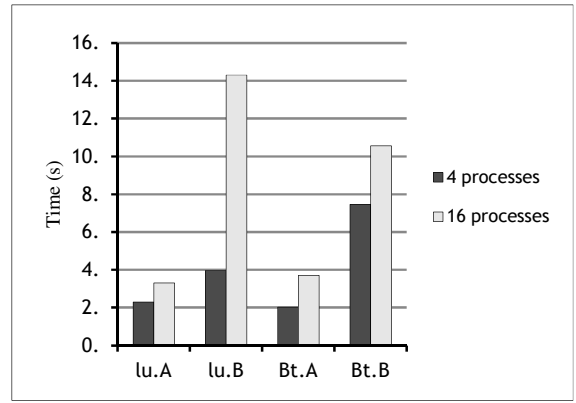

Fig. 8. Single Amazon EC2 node restart time (DMTCP).

The next set of experiments evaluated the performance of the selected checkpoint/restart tools while increasing the number of computing processes on Amazon EC2. We compiled the used NAS parallel benchmarks for 16 processes and for the selected classes. The DMTCP tool was selected for this scalability test in case of running the newly compiled benchmarks on a single node on Amazon EC2, with overloaded 16 processes. So on this single Amazon EC2 instance, same tests were repeated for both 4 and 16 processes using DMTCP as its checkpoint/restart technique. The DMTCP checkpoint time and restart time are shown in Figure 7 and Figure 8 respectively. On the other hand, BLCR integrated into the OpenMPI framework was tested using 4-compute nodes cluster on Amazon EC2, with overloaded 16 processes (4 processes per instance). On this 4 Amazon EC2 instances cluster, same tests were repeated for both 4 and 16 processes using BLCR integrated into OpenMPI as its checkpoint/restart technique. The BLCR checkpoint time and restart time are shown in Figure 9 and Figure 10 respectively. Summarizing all experiments to chart the data scalability and the participating processes scalability trends are shown in Figures 11 and 12, with an exponential trend line in both cases. Tables for all experiments' results are published in http://www.manalhelal.com/publications/cr-comparisons/.

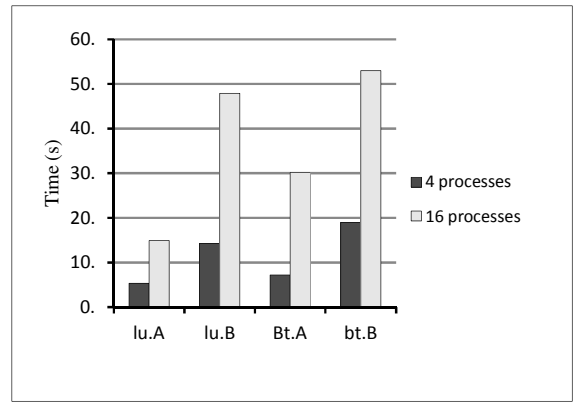

Fig. 9. 4-nodes Amazon EC2 cluster checkpoint time (BLCR)

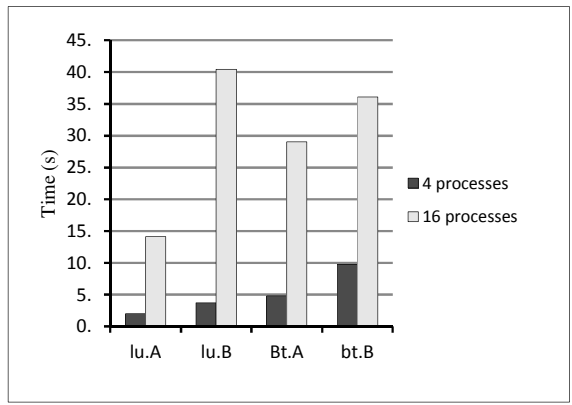

Fig. 10. 4-nodes Amazon EC2 cluster restart time (BLCR)

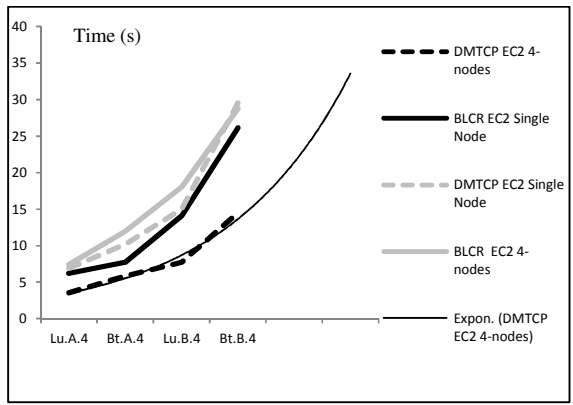

Fig. 11. Data scalability trend chart

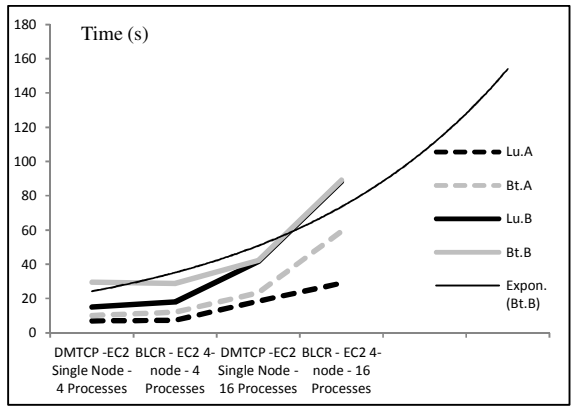

Fig. 12. Processes scalability trend chart

## VI. ANALYSIS AND CONCLUSIONS

Migrating distributed parallel applications to the cloud has significantly increased in the last year. This requires ensuring that parallel applications that use a legacy tool as a fault tolerance technique can work properly also on the cloud. It has been realised from the results of the experiments done in this work that two of the most commonly used checkpoint/restart tools, which are DMTCP and BLCR integrated into OpenMPI can perform well on Amazon EC2 either on a single node or on a cluster of a number of computing nodes. Figure 1, 2, 3, 4, 5 and 6 shows that DMTCP is faster than BLCR integrated into OpenMPI. DMTCP has the lower checkpoint and restart time values.

Some overhead reasons which are related to checkpoint I/O using BLCR integrated into OpenMPI explains why it acts slower than DMTCP. To checkpoint an application there is a sheer amount of data to be stored to a persistent storage. Excessive input/output operations (I/O) that may occur to access these saved checkpoint files can negatively affect the application performance due to several overheads [42, 43]. It is noticed that to take a snapshot with BLCR integrated into MPI, a large number of ineffective and relatively small write operations have to be done to save the checkpoint files. This results in degradation in the underlying file system performance. It becomes more susceptible to have many severe I/O contentions because of these simultaneous write accesses. These unfavourable consequences can possibly occur at intra-node level in case of utilising multi-core architecture hosting a number of concurrent processes on one node and also at inter-node level where several processes from different nodes within the cluster context are simultaneously executed. The IO contentions decrease IO throughput. It also leads to a large variation for each individual process to complete its checkpoint writing. So if some of the participating processes take a lower time to finish their checkpoint writing, they should wait till the slower processes finish writing their snapshots to coordinate with each other to have a global consistent checkpoint. Then they can all resume the execution [44].

DMTCP tends to be more conservative at checkpoint and restart times. It avoids situations in which performance degradation may occur. For Example, The DMTCP library is designed to be aware of all necessary information required to perform checkpoint and restart of the distributed application. So DMTCP knows all the details about forked child processes, created remote processes and created sockets parameters. For this purpose DMTCP adds wrappers around a number of libc functions by overriding libc symbols with DMTCP library. To obtain efficient performance, it is strictly avoided to place wrappers around any frequently invoked system calls such as read () and write () because it may produce measurable overhead [8].

Both DMTCP and BLCR integrated into OpenMPI produce a checkpoint image file in a local partition per each node. In case of BLCR, the checkpoint/restart service of OpenMPI write the checkpoint to a local directory, and then copies each local checkpoint to a central coordinator process located at a single central node. This results in a serialization of a part of the parallel checkpoint production consuming more time to write and read checkpoint images when performing checkpoint or restart, so it slows the performance of OpenMPI with BLCR Support [3,7,8].

In the section of experiment in which the two selected techniques were tested on a single Amazon EC2 instance, DMTCP was reconfigured and reinstalled to enable an additional facility called Forked Checkpoint. DMTCP documentation claimed that this option improves checkpoint speed by forking a child process to write the checkpoint image while the parent process continues the application execution. The experiment results shown in figure 3 show that DMTCP configured to use forked checkpointing is still faster than OpenMPI with BLCR Support. It was advised by DMTCP documentation to test whether the forked checkpointing optimization is compatible with user applications. So it was preferred to test it once and continue testing DMTCP using the default configuration in the rest of experiments in order to have results that will be valid for the vast majority of user's applications [8].

One of the aims of scalability tests concerning distributed processing is to focus on computations consisting of many processes. Tests we performed on a single Amazon EC2 node or an Amazon EC2 cluster of 4-computing nodes indicate that for both tools there is an increase in the checkpoint and restart time values when increasing the number of computation processes as shown in Figure 7, 8, 9 and 10. The figures indicate that DMTCP has a higher potential to scale well to a larger number of computation processes than BLCR integrated into the OpenMPI framework. Differences between values of checkpoint time for the same application in the same environment using 4 and 16 processes are larger in case of OpenMPI with BLCR support. The same notice was found for restart time values. It is noticed that BLCR in average takes 3.28 more time for 16 processes than for 4 processes in checkpointing. While DMTCP in average takes 2.5 more time for 16 processes than for 4 processes. The limited time and budget available for these experiments prevented repeating the experiments to fix all errors and reach a totally unified environment. Therefore the fact that the BLCR experiments used 16 processes on 4 nodes adds a justified communication delay that does not exist in the 16 processes in a single node in the DMTCP experiment. However, the previously mentioned overhead reasons associated with the way in which BLCR writes processes snapshots is participating in the abrupt increase in checkpoint and restart time values when increasing the number of computation processes. The checkpoint files in case of 16 processes will have more data and will be of bigger sizes so the performance will be slower by a considerable rate.

Data scalability and the compute processes scalability trends are shown in Figures 11 and 12, with an exponential trend line in both cases.

## VII. FUTURE WORK

A lot of research has been accomplished on checkpoint/restart. However several factors are not yet addressed. Concerning the presented experiment, higher research budget will enable further testing to be conducted on the selected tools to test their performance in response to changing factors. These factors include changing the testing environment. One example is adding more nodes virtualised in the same physical machine vs. nodes that are close geographically vs. nodes that are dispersed geographically. Other examples include using different network technologies, and different cloud providers such as Microsoft Azure. The experiment design itself can include other compute intensive applications vs. communication intensive applications. Furthermore, other tools can be added to enrich the assessment process.

REFERENCES

[1] I. Foster, Y. Zhao, I. Raicu, and S. Lu, "Cloud computing and grid computing 360-degree compared" In Proc. IEEE Grid Computing Environments Workshop, IEEE Press, 2008, pp. 1-10.

[2] W. Gropp, E. Lusk, and R. Thakur, "Using MPI-2: advanced features of the message passing interface" MIT Press, Vol. 1, 1999.

[3] J. Hursey, J. M. Squyres, T. I. Mattox, and A. Lumsdaine, "The design and implementation of checkpoint/restart process fault tolerance for Open MPI" In Proc. 21st IEEE International Parallel and Distributed Processing Symposium (IPDPS). IEEE Computer Society, 2007.

[4] E. N. M. Elnozahy, L. Alvisi, Y.-M. Wang, and D. B. Johnson, "A survey of rollback-recovery protocols in message passing Systems" ACM Computing Surveys, Vol. 34, No. 3, 2002, pp. 375-408.

[5] J. Duell, P. Hargrove, and E. Roman, "The design and implementation of berkeley lab's linux checkpoint/restart" Lawrence Berkeley National Laboratory Technical Report, 2002

[6] P. H. Hargrove, and J. C. Duell, "Berkeley lab checkpoint/restart (BLCR) for linux clusters" Journal of Physics: Conference Series, Vol. 46, No. 1. IOP Publishing, 2006, pp. 494-499.

[7] J. Hursey, J. M. Squyres, and A. Lumsdaine, "A checkpoint and restart service specification for Open MPI" Technical Report, Indiana University, Bloomington, Indiana, USA, Tech. Rep. TR635, 2006.

[8] J. Ansel, K. Arya, and G. Cooperman, "DMTCP: transparent checkpointing for cluster computations and the desktop" in Proc. Int. Parallel and Distributed Processing Symposium (IPDPS'09), IEEE, 2009.

[9] K. Jackson, L. Ramakrishnan, K. Muriki, S. Canon, S. Cholia, J. Shalf, H. Wasserman, and N. Wright, "Performance analysis of high performance computing applications on the amazon web services cloud" 2nd IEEE Int. Conf. on Cloud Computing Technology and Science. IEEE, 2010, pp. 159-168.

[10] S. Yi, A. Andrzejak, and D. Kondo, "Monetary cost-aware checkpointing and migration on amazon cloud spot instances" IEEE Transactions on Services Computing, 2012.

[11] R. K. Sahoo, R. K., A. Sivasubramaniam, M. S. Squillante, and Y. Zhang, "Failure data analysis of a large-scale heterogeneous server environment" In Proc. Int. Conf. on Dependable Systems and Networks (DSN'04), Florence, Italy, 2004, pp. 772-781.

[12] A. Bala, and I. Chana, "Fault tolerance-challenges, techniques and implementation in cloud computing" International Journal of Computer Science Issues (IJCSI), Vol. 9, No. 1, 2012.

[13] G. Bosilca, A. Bouteiller, F. Cappello, S. Djilali, G. Fedak, C. Germain, T. Herault, P. Lemarinier, O. Lodygensky, F. Magniette, V. Neri, and A. Selikhov, "Mpich-v: toward a scalable fault tolerant MPI for volatile nodes" High Performance Networking and Computing (SC2002), USA, 2002, IEEE/ACM.

[14] I. P. Egwutuoha, D. Levy, B. Selic, and S. Chen, "A survey of fault tolerance mechanisms and checkpoint/restart implementations for high performance computing systems" The Journal of Supercomputing, Vol. 65, No. 3, 2013, pp. 1302-1326.

[15] A. Maloney, and A. Goscinski, "A survey and review of the current state of rollback-recovery for cluster systems" Concurr. Comput. Pract. Exper. , Vol. 21, No. 12, 2009, pp. 1632-1666.

[16] A. Nagarajan, F. Mueller, C. Engelmann, and S. Scott, "Proactive fault tolerance for HPC with Xen virtualization" 21st annual int. conf. on Supercomputing, Seattle, USA, 2007, pp. 23-32.

[17] Checkpointing.org, The Home to Checkpointing Packages. (http://checkpointing.org). N.p., n.d. Web. 5 May. 2013.

[18] S. Kalaiselvi, and V. Rajaraman, "A survey of checkpointing algorithms for parallel and distributed computers" Sadhana, Vol. 25, No. 5, 2000, pp. 489-510.

[19] B.-J. Kim, "Comparison of the existing checkpoint systems" Technical report, IBM Watson, 2005.

[20] E. Roman, "A survey of checkpoint/restart implementations" Lawrence Berkeley National Laboratory Technical Report, Tech. Rep. LBNL-54942, 2002.

[21] J. Duell, P. Hargrove, and E. Roman, "Requirements for linux checkpoint/restart" Lawrence Berkeley National Laboratory Technical Report, Tech. Rep. LBNL-49659, 2002.

[22] K. M. Chandy, and L. Lamport, "Distributed snapshots: determining global states of distributed systems" ACM Transactions on Computing Systems, Vol. 3, No. 1, 1985, pp. 63-75.

[23] E. N. Elnozahy, D. B. Johnson, and W. Zwaenepoel, "The performance of consistent checkpointing" In Proc. 11th Symposium on Reliable Distributed Systems, 1992, pp. 39-47.

[24] S. Sankaran, J. M. Squyres, B. Barrett, A. Lumsdaine, J. Duell, P. Hargrove, and E. Roman, "The LAM/MPI checkpoint/restart framework: system-initiated checkpointing" International Journal of High Performance Computing Applications, Vol. 19, No. 4, 2005, pp. 479-493.

[25] S. Sankaran, J. M. Squyres, B. Barrett, A. Lumsdaine, J. Duell, P. Hargrove, and E. Roman, "Parallel checkpoint/restart for MPI applications" International Journal of High Performance Computing Applications, Vol. 1, No. 4, 2005, pp. 479-493.

[26] C. E. Leiserson, Z. S. Abuhamdeh, D. C. Douglas, C. R. Feynman, M. N. Ganmukhi, J. V. Hill, W. D. Hillis, B. C. Kuszmaul, M. A. St Pierre, D. S. Wells, M. C. Wong-Chan, S. Yang, and R. Zak, "The network architecture of the connection machine CM-5" Journal of Parallel and Distributed Computing, Vol. 33, No. 2, 1996, pp. 145-158.

[27] G. Zheng, L. Shi, and L. V. Kale, "FTC-Charm++: An in-memory checkpoint-based fault tolerant runtime for charm++ and MPI" in Proc. Int. Conf. on Cluster Computing. IEEE, 2004, pp. 93-103.

[28] C. Wang, F. Mueller, C. Engelmann, and S. L. Scott, "Hybrid full/incremental check-point/restart for MPI jobs in HPC environments" In Proc. Int. Conf. on Parallel and Distributed Systems, 2011.

[29] B. Randell, "System structure for software fault tolerance" IEEE Transactions on Software Engineering, Vol. SE-1, No. 10, 1975, pp. 1220-232.

[30] A. Guermouche, T. Ropars, E. Brunet, M. Snir, and F. Cappell, "Uncoordinated checkpointing without domino effect for send-deterministic message passing applications" 25th IEEE International Parallel & Distributed Processing Symposium (IPDPS2011), Anchorage, USA, 2011.

[31] K. Arya, G. Cooperman, A. Dotti, and P. Elmer, "Use of checkpoint-restart for complex HEP software on traditional architectures and Intel MIC" Journal of Physics: Conference Series 523, Vol. 523, No. 1. IOP Publishing, 2014.

[32] M. Rieker, J. Ansel, and G. Cooperman, "Transparent user-level checkpointing for the native POSIX thread library for linux" In Proc. Parallel and Distributed Processing Techniques and Applications (PDPTA-06), 2006, pp. 492-498.

[33] D. K. Panda, "MVAPICH2: A High performance MPI library for NVIDIA GPU clusters with InfiniBand " In Proc. GPU TechnologyConference (GTC2013), San Jose, 2013.

[34] AWS | Amazon EC2 | Instance Types, Amazon Web Services, Inc. (http://aws.ama-zon.com/ec2/instance-types/). N.p., n.d. Web. 14 Sep. 2013.

[35] Amazon's Physical Hardware and EC2 Compute unit, HuanLius Blog. (http://huanliu.wordpress.co-m/2010/06/14/amazons-physical-hardware-and-ec2-compute-unit/). N.p., n.d. Web. 13 Sep. 2013.

[36] StarCluster, STAR: Cluster. (http://star.mit.edu/cluster/). N.p., n.d. Web. 1 Mar. 2014.

[37] AWS | Amazon Elastic Block Store (EBS) – Persistent Storage, (http://aws.amazon.com/ebs/). N.p., n.d. Web. 1 June. 2014.

[38] AWS | Amazon Simple Storage Service (S3) – Cloud Storage, (http://aws.amazon.com/s3/). N.p., n.d. Web. 1 June. 2014.

[39] F. Wong, R. Martin, R. Arpaciusseau, and D.Culler, "Architectural requirements and scalability of the NAS parallel benchmarks" In Proc. Supercomputing '99, Portland, OR, 1999.

[40] H. Jung, H. Han, H. Y. Yeom, and S. Kang, "Athanasia: a user-transparent and fault-tolerant system for parallel applications" IEEE Trans. Parallel Distrib. Syst. Vol.22, No.10, 2011, pp. 1653-1668.

[41] NASA Advanced Supercomputing Division, NAS Parallel Benchmark (http://www.nas.nasa.gov/publications/npb.html ). N.p., n.d. Web. 17 Mar. 2013.

[42] J. S. Plank, Y. Chen, K. Li, M. Beck, and G. Kingsley, "Memory exclusion: Optimizing the performance of checkpointing systems" SW: Practice and Experience, Vol.29, No.2, 1999, pp. 125-142.

[43] J. Bent, G. Gibson, G. Grider, B. McClelland, P. Nowoczynski, J. Nunez, M. Polte, and M. Wingate, "PLFS: a checkpoint filesystem for parallel applications" In Proc. ACM/IEEE Transactions on Computing Conference on High Performance Networking and Computing (SC '09), Portland OR, 2009.

[44] X. Ouyang, R. Rajachandrasekar, X. Besseron, H. Wang, J. Huang, and D. K. Panda, "CRFS: A lightweight user-level filesystem for generic checkpoint/restart" in Proc. Int. Conf. on Parallel Processing, ser. ICPP '11. Washington, DC, USA: IEEE Computer Society, 2011, pp. 375-384.